\documentclass[twocolumn,epjc3]{svjour3}
\usepackage{amsmath,amssymb}
\usepackage{graphicx}
\graphicspath{{figures/}}
\usepackage{hepnames}
\usepackage{slashed}
\usepackage{siunitx}
\usepackage{multirow}
\usepackage{mathptmx}
\usepackage{latexsym}
\usepackage{flushend}
\usepackage[numbers,sort&compress]{natbib}
\usepackage[colorlinks,citecolor=blue,urlcolor=blue,linkcolor=blue]{hyperref}

\begin{document}

\title{Finite-volume effects on phase transition in the Polyakov-loop
extended Nambu-Jona-Lasinio model with a chiral chemical potential}

\author{Zan Pan\thanksref{e1,addr1,addr2}
 \and Zhu-Fang Cui\thanksref{e2,addr3}
 \and Chao-Hsi Chang\thanksref{e3,addr1,addr2,addr4}
 \and Hong-Shi Zong\thanksref{e4,addr1,addr3,addr5}}
\thankstext{e1}{e-mail: panzan@itp.ac.cn}
\thankstext{e2}{e-mail: phycui@nju.edu.cn}
\thankstext{e3}{e-mail: zhangzx@itp.ac.cn}
\thankstext{e4}{e-mail: zonghs@nju.edu.cn}
\institute{Key Laboratory of Theoretical Physics,
 Institute of Theoretical Physics, CAS, Beijing, 100190, China \label{addr1}
 \and
 School of Physical Sciences, University of Chinese Academy of Sciences,
 Beijing 100049, China \label{addr2}
 \and
 Department of Physics, Nanjing University,
 Nanjing 210093, China \label{addr3}
 \and
 CCAST (World Laboratory), P.O. Box 8730,
 Beijing 100190, China \label{addr4}
 \and
 Joint Center for Particle, Nuclear Physics and Cosmology,
 Nanjing 210093, China \label{addr5}
}

\date{Received: \today}

\maketitle

\begin{abstract}
To investigate finite-volume effects on the chiral symmetry restoration
and the deconfinement transition and some impacts of possible global topological
background for a quantum chromodynamics (QCD) system with $N_f=2$ (two
quark flavors), we apply the Polyakov-loop extended Nambu-Jona-Lasinio model by introducing
a chiral chemical potential $\mu_5$ artificially. The final numerical results indicate
that the introduced chiral chemical potential does not change the critical exponents
but shifts the location of critical end point (CEP) significantly;
the ratios for the chiral chemical potentials and temperatures at CEP,
$\mu_c/\mu_{5c}$ and $T_c/T_{5c}$, are significantly affected by the system size $R$.
The behavior is that $T_c$ increases slowly with $\mu_5$ when $R$ is `large'
and $T_c$ decreases first and then increases with $\mu_5$ when $R$ is `small'.
It is also found that for a fixed $\mu_5$, there is a $R_{\text{min}}$,
where the critical end point vanishes, and the whole phase diagram becomes a
crossover when $R<R_{\text{min}}$. Therefore, we suggest that
for the heavy-ion collision experiments, which is to study the possible
location of CEP, the finite-volume behavior should be taken into account.
\end{abstract}

\section{Introduction}

The thermodynamics of strongly interacting matter ruled by 
quantum chromodynamics (QCD) under extreme conditions of
temperature and density is a profound and challenging area overlapping
statistical, particle and nuclear physics. A deep understanding of its phase structure
of QCD is expected to bring some insights on many fundamental problems such as
compact stars, and the early universe~\cite{Rajagopal2000, Satz2012}.
Experiments with heavy-ion collisions such as the BNL Relativistic Heavy-Ion Collider
(RHIC) and the CERN Large Hadron Collider (LHC) are continuing active
investigations on the strongly interacting matter in the
laboratory~\cite{PhysRevLett.112.032302, Abelev2014139}.

It is expected that QCD could lead to a rich phase
structure of matter~\cite{Fukushima2011}. Lattice simulations from the first principle
have revealed that the confined quarks will become released to quark-gluon plasma
around the temperature $T_c=\SI{154\pm 9}{MeV}$~\cite{PhysRevD.85.054503, Petreczky2012}.
However, due to the sign problem, Monte Carlo methods can
be only directly applied to states around zero quark density.
Therefore, effective models which exhibit the features of color confinement
and spontaneous chiral symmetry breaking are more feasible to be used to
study the phase structure of QCD on the density in addition to temperature.
Here, we will adopt the Polyakov-loop extended 
Nambu-Jona-Lasinio (PNJL) model with chiral and axial chemical
potentials~\cite{Fukushima2004277, PhysRevD.73.014019, PhysRevD.82.076003, PhysRevD.85.054013}.

In the PNJL model with some assumptions, it is found that
the chiral symmetry restoration and deconfinement transition may coincide,
i.e. discontinuities appear simultaneously in their order parameters,
the chiral condensate $\sigma$ and the Polyakov loop $L$, and they are of the first 
order~\cite{Fukushima2004277, PhysRevD.73.014019, PhysRevD.82.076003, PhysRevD.85.054013}.
A nontrivial critical end point (CEP) also exists at finite temperature $T$ 
and a quark chemical potential $\mu$, which indicates a coincidence of 
second-order phase transitions~\cite{Fukushima2004277, PhysRevD.73.014019, 
PhysRevD.82.076003, PhysRevD.85.054013}. Although the axial currents 
are not conserved due the global topological solutions of QCD (such as instantons etc.) 
and quantum anomaly~\cite{PhysRevD.78.074033}, to characterize imbalance on 
the chirality in terms of $N_L$ and $N_R$ ($N_R+N_L=N$, $N_R-N_L=N_5$) for the quark matter,
we follow the authors of Ref.~\cite{PhysRevD.78.074033}, similar to 
the chemical potential $\mu$, introduce a chiral chemical potential $\mu_5$
(conjugate to $N_5$) as a mathematical artifice to mimic the effects of
`chiral transitions'. This also leads to a novel idea, proposed
in Ref.~\cite{PhysRevD.84.014011}, to detect the CEP by simulating QCD but
with the chiral chemical potential $\mu_5$, then by continuation in the
plane of $\mu_5$-$\mu$, the critical end point CEP denoted by $\text{CEP}_5$ can be obtained,
which is accessible to lattice QCD simulations of grand-canonical ensembles.
Thus the authors of Ref.~\cite{PhysRevD.84.014011} believe that it should be a signal for the
existence of the CEP in QCD if a $\text{CEP}_5$ is found by lattice simulation. Besides,
since physical systems in the same universality class all share the
same critical exponents, so here we would like to check if this
continuation is reasonable by precise checking the impacts on the critical exponents
brought by the introduction of $\mu_5$.

For the two-flavor PNJL model, there are unphysical decays of
hadrons to quarks due to the spurious poles in the quark
loop diagrams~\cite{PhysRevD.75.065004, PhysRevD.91.125040}.
Introducing a lower momentum cutoff to mimic confining effects of
strong interaction helps to address this problem.
This is also the starting point for us further to incorporate the finite-volume effects.
In experiments of heavy-ion collisions, the strongly interacting matter
formed through the energy deposition of the colliding particle definitely
has a finite volume. Therefore, it is very important to have a knowledge
on the finite-volume effects when studying the thermodynamic phases experimentally.
In the context of heavy-ion collisions,
the importance of finite-volume effects in the thermodynamics of
strong interaction matter may be brought forward with the help of finite size
scaling analysis~\cite{Palhares2011, PhysRevC.84.011903}. In the past years,
many theoretical studies of finite-volume effects have been performed on
NJL models~\cite{Abreu2006551, PhysRevD.74.054036, PhysRevD.83.025001}.
However, only recently the studies on the thermodynamic properties of
strongly interacting matter in a finite volume using the PNJL models
have aroused increasing attention~\cite{PhysRevD.81.114017,
PhysRevD.87.054009, PhysRevD.91.051501}.

Our paper is organized as follows. First, in Sec.~\ref{sec:pnjl} we briefly review
the PNJL model with chemical and chiral chemical potentials in the mean field approximation.
In Sec.~\ref{sec:phases}, numerical results on chiral symmetry restoration
and deconfinement transition for $N_f=2$ model with the infinite size are presented.
We also verify that the chiral chemical potential does not impact on the value of
critical exponents. By introducing the lower momentum cutoff, we investigate
its finite-volume effects in Sec.~\ref{sec:volume}. Finally,
in Sec.~\ref{sec:conclusions} we summarize our results and some conclusions
are made.

\section{\label{sec:pnjl}The PNJL model and its extension in mean field
approximation}

In this section, firstly we briefly review the PNJL model and its extension 
in the mean field approximation~\cite{PhysRevD.85.054013, PhysRevD.73.014019}. 
The Lagrangian for PNJL model is given by
\begin{equation}
 \label{eq:lagrangian}
 \mathcal{L}=\bar{\psi}(i\slashed{D}-m)\psi
 +G\left[(\bar{\psi}\psi)^2+(i\bar{\psi}\gamma_5\bm{\tau}\psi)^2\right]
 -\mathcal{U}(L,L^{\dag},T),
\end{equation}
where $\psi=(u,d)$ represents the quark fields; so the number of flavors
is taken as $N_f=2$, and the number of colors is $N_c=3$;
the two-flavor current quark mass matrix is $m=\mathrm{diag}(m_u,m_d)$,
and we shall work in the isospin-symmetric limit with $m_u=m_d$;
$\bm{\tau}$ corresponds to the Pauli matrices in flavor space.

The potential term $\mathcal{U}(L,L^{\dag},T)$ is the effective potential
expressed in terms of the traced Polyakov loop $L$ and its conjugate
\begin{equation}
 L=\frac{1}{N_c}\mathrm{Tr}_c W,\quad
 L^{\dag}=\frac{1}{N_c}\mathrm{Tr}_c W^{\dag}.
\end{equation}
The Polyakov loop $W$ is a matrix in color space explicitly given by
\begin{equation}
 W=\mathcal{P}\exp\bigg[i\int_0^{\beta}A_4(\bm{x},\tau)\,\mathrm{d}\tau\bigg],
\end{equation}
where $\beta=1/T$ is the inverse temperature and $A_4=iA^0$.
In the Polyakov gauge, $W$ can have a diagonal representation
in color space~\cite{Fukushima2004277}. The traced Polyakov loop
$L$ is an exact order parameter of spontaneous $\mathbb{Z}_3$ symmetry
breaking in pure gauge theory. Although in full QCD the presence of
dynamical quarks explicitly breaks the $\mathbb{Z}_3$ symmetry,
it still seems to be a good indicator of the deconfinement phase transition.
To incorporate the confinement or deconfinement properties,
we have introduced a Polyakov-loop-dependent coupling constant $G$ as
\begin{equation}
 G=g\big[1-\alpha_1LL^{\dag}-\alpha_2(L^3+L^{\dag 3})\big].
\end{equation}
For simplicity we will take $L=L^{\dag}$
following Refs.~\cite{PhysRevD.85.054013, PhysRevD.84.014011},
which implies $A_4^8=0$. As shown in Ref.~\cite{Rossner2008118},
the effects of fluctuations leading to $L\ne\bar{L}$ turn out not to be of major
qualitative importance in determining the phase diagram.
The numerical values of $\alpha_1$ and $\alpha_2$ can be obtained by
a best fit of lattice data at zero and imaginary chemical potential,
which leads to $\alpha_1=\alpha_2=0.2$.

Now let us consider the extended PNJL model, since we would like to treat quark
matter with chirality imbalance $N_5=N_R-N_L$. The way to do it is, besides the
chemical potential $\mu$ conjugated to the density $n$, additionally
to introduce a chiral chemical potential $\mu_5$ conjugated to the
chiral density $n_5$~\cite{PhysRevD.81.114031, PhysRevD.83.105008, PhysRevD.84.014011}.
At Lagrangian level, this amounts to add the density operator
$\mu\bar{\psi}\gamma^0\psi$ and the chiral density operator
$\mu_5\bar{\psi}\gamma^0\gamma^5\psi$ to Eq.~\eqref{eq:lagrangian}.
Namely the Lagrangian~\eqref{eq:lagrangian} for the model has been extended to the following
\begin{equation}
 \label{eq:chiral-lagrangian}
 \begin{split}
  \mathcal{L}&=\bar{\psi}(i\slashed{D}-m+\mu\gamma^0+\mu_5\gamma^0\gamma^5)\psi \\
  &\qquad +G\left[(\bar{\psi}\psi)^2+(i\bar{\psi}\gamma_5\bm{\tau}\psi)^2\right]
   -\mathcal{U}(L,L^{\dag},T).
 \end{split}
\end{equation}

Making the mean field approximation and performing the path integral
over the quark field, we can obtain the thermodynamic potential density
$\mathcal{V}$ at the one-loop level
\begin{equation}
 \label{eq:thermodynamic-potential}
 \begin{split}
  \mathcal{V}&=\mathcal{U}(L,L^{\dag},T)+G\sigma^2
   -N_cN_f\sum_{s=\pm 1}\int\frac{\mathrm{d}^3\bm{p}}{(2\pi)^3}\,\omega_s \\
  &\qquad -N_f\sum_{s=\pm 1}\int\frac{\mathrm{d}^3\bm{p}}{(2\pi)^3}\,T\log(F_{+}F_{-}),
 \end{split}
\end{equation}
where $\sigma=\langle\bar{\psi}\psi\rangle$ is the chiral condensate
and relates to the effective quark mass $M$ as
\begin{equation}
 M=m-2G\sigma.
\end{equation}
The index $s$ denotes the helicity projection and
\begin{equation}
 \omega_s=\sqrt{(|\bm{p}|s-\mu_5)^2+M^2}
\end{equation}
corresponds to the pole of the quark propagator. The momentum integral
for $\omega_s$ corresponds to the vacuum quark fluctuations and
a momentum cutoff $\Lambda$ is introduced to regularize it here.

The last term in Eq.~\eqref{eq:thermodynamic-potential} is responsible for
the statistical properties of the model at low temperature.
Therein we have introduced the functions
\begin{equation}
 \begin{split}
  F_{-}&=1+3Le^{-\beta(\omega_s-\mu)}
   +3L^{\dag}e^{-2\beta(\omega_s-\mu)}+e^{-3\beta(\omega_s-\mu)}, \\
  F_{+}&=1+3L^{\dag}e^{-\beta(\omega_s+\mu)}
   +3Le^{-2\beta(\omega_s+\mu)}+e^{-3\beta(\omega_s+\mu)}.
 \end{split}
\end{equation}
In order to reproduce the pure gluonic lattice data
with $N_c=3$, the potential term $\mathcal{U}$ is taken as
the following form
\begin{equation}
 \begin{split}
  \mathcal{U}(L,L^{\dag},T)&=T^4\bigg\{{-\frac{1}{2}}a(T)LL^{\dag}
    +b(T)\ln\big[1-6LL^{\dag} \\
   &\qquad +4(L^3+L^{\dag 3})-3(LL^{\dag})^2\big]\bigg\},
 \end{split}
\end{equation}
where the model parameters are given by
\begin{align}
 a(T)&=a_0+a_1\left(\frac{T_0}{T}\right)+a_2\left(\frac{T_0}{T}\right)^2, \\
 b(T)&=b_3\left(\frac{T_0}{T}\right)^3.
\end{align}
The choice of coefficients from the mean field approximation reads
\begin{equation}
 a_0=3.51,\quad a_1=-2.47,\quad a_2=15.2,\quad b_3=-1.75.
\end{equation}
Other numerical parameters used in our calculations are taken
as those in Ref.~\cite{PhysRevD.85.054013}
\begin{equation}
 \begin{split}
  T_0&=\SI{190}{MeV},\quad \Lambda=\SI{631.5}{MeV},\\
  m&=\SI{5.5}{MeV},\quad g=\SI{5.498e-6}{MeV^{-2}}.
 \end{split}
\end{equation}

\section{\label{sec:phases}Chiral symmetry restoration and deconfinement transition 
in the extended PNJL model}

In this section we will present the way to find chiral symmetry restoration and deconfinement
transition in the PNJL model with chemical potential $\mu$ and chiral chemical potential $\mu_5$,
and show its reasonableness for investigating the QCD phase transitions.

For any given $\mu$, $\mu_5$ and $T$, we can obtain the corresponding values of
$\sigma$ and $L$ by solving the gap equations
\begin{equation}
 \label{eq:gap-equations}
 \frac{\partial\mathcal{V}}{\partial\sigma}=0,\quad
 \frac{\partial\mathcal{V}}{\partial L}=0.
\end{equation}

However, this approach is hard to work in practice due to the difficulty
in solving the coupled integral equations by means of iterative methods.
Moreover, the solutions of these equations do not necessarily yield
a global minimum. There are possibilities that they may yield a local minimum,
even a maximum. Therefore here not only to find the solutions of Eq.~\eqref{eq:gap-equations}
we also need to check that the corresponding solution yields a global minimum
when they are inserted back into Eq.~\eqref{eq:thermodynamic-potential}.
For a better approach, we can solve the problem in another way, i.e.
to solve Eq.~\eqref{eq:gap-equations} by finding the minima of
the potential function $\mathcal{V}$. This reduces to the famous problem of
multidimensional minimization, and here we can use the efficient
Nelder-Mead simplex algorithm.

\begin{figure}
 \includegraphics[width=235pt]{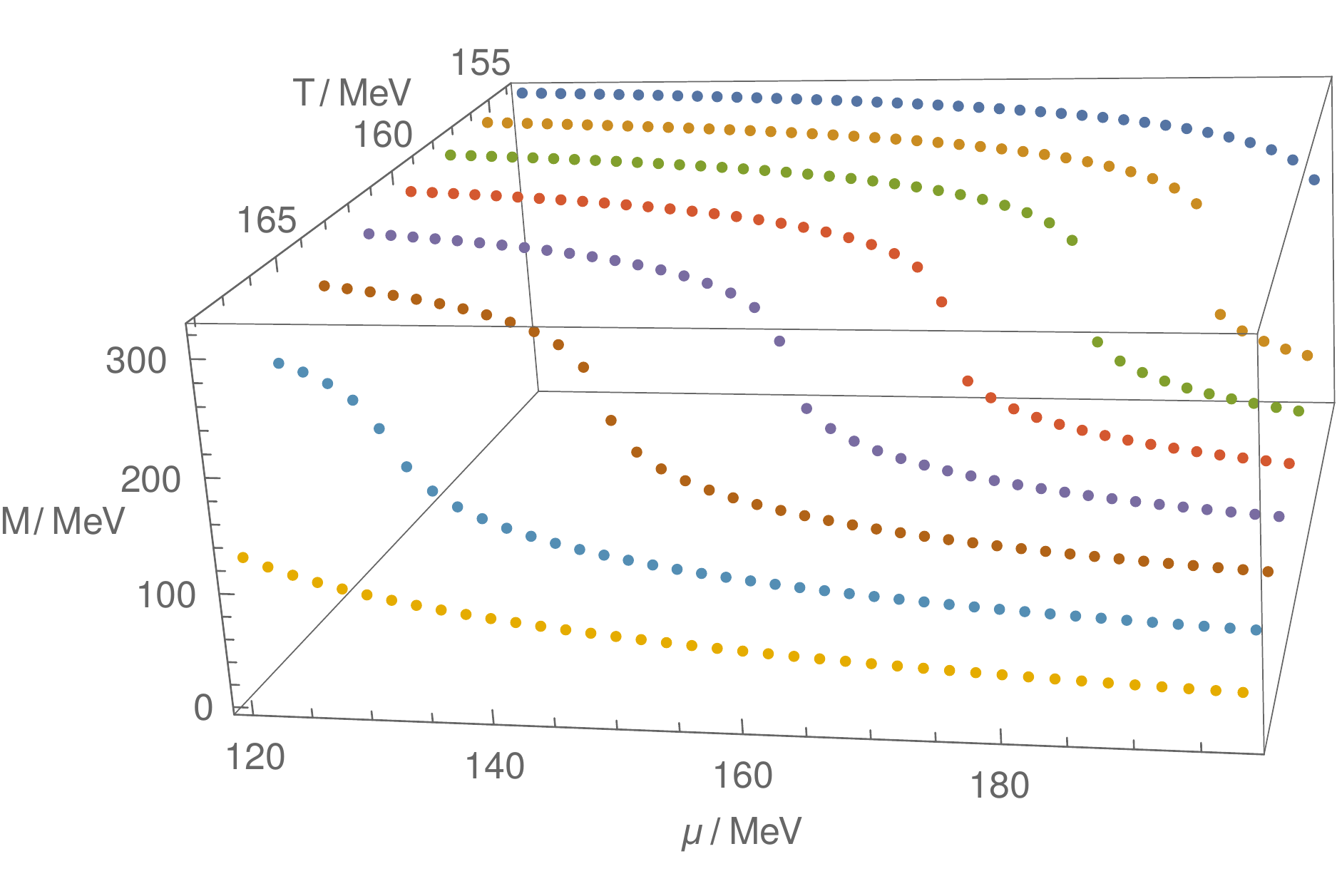}
 \includegraphics[width=235pt]{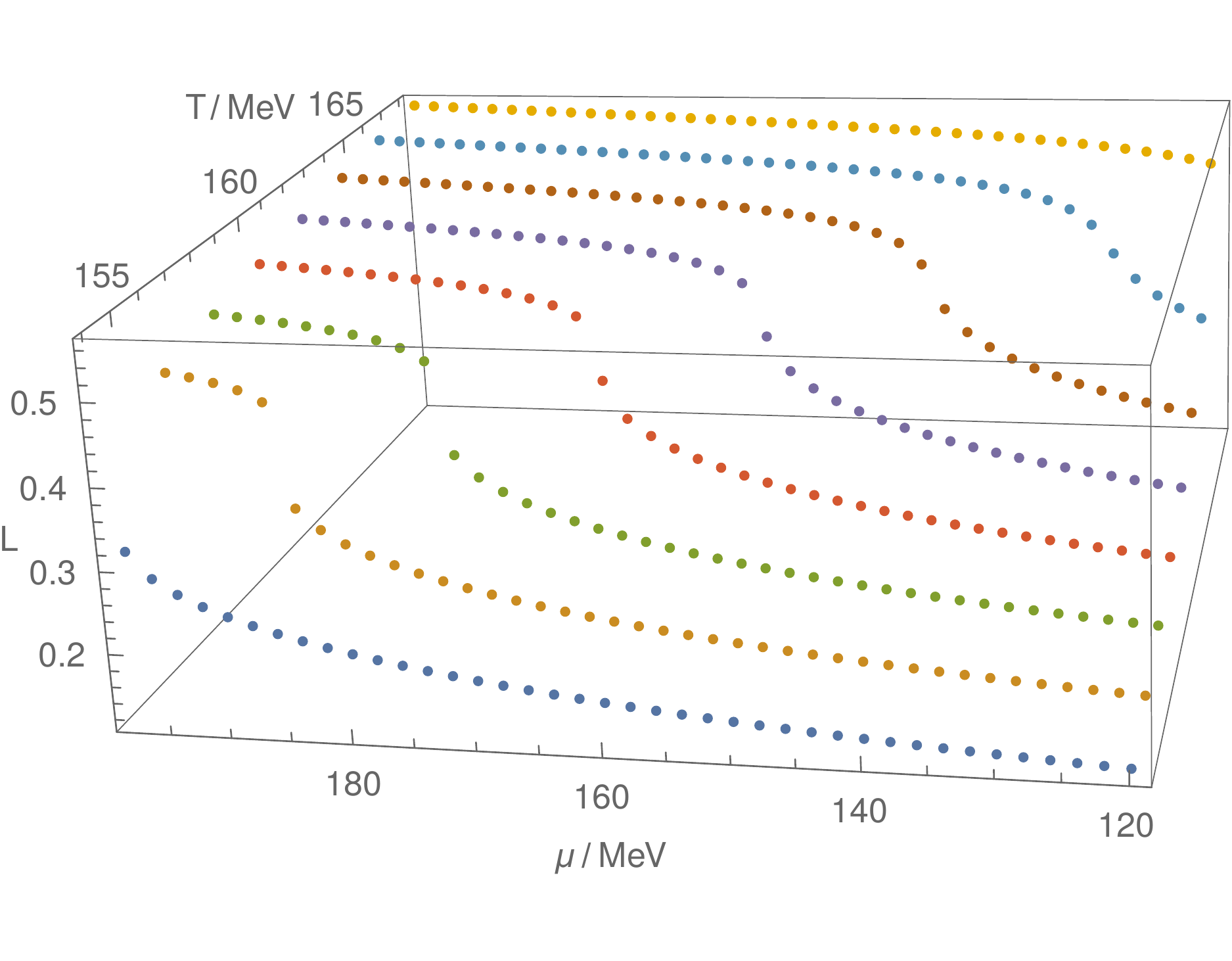}
 \caption{3D plot for the effective mass $M$ (upper panel) in the $\mu$-$T$-$M$ space
  and the Polyakov loop $L$ (lower panel) in the $\mu$-$T$-$L$ space
  near the CEP $(\mu_c, T_c)=(172.7, 159.2)$, where $\mu_5=0$.}
 \label{fig:mu5-zero-3d}
\end{figure}

First, we consider the case when $\mu_5=0$. For any given $\mu$ and T,
we can calculate the solutions of Eq.~\eqref{eq:gap-equations}.
As shown in Fig.~\ref{fig:mu5-zero-3d}, the discontinuity of the effective
mass $M$ and that of the Polyakov loop $L$ vanish simultaneously at the same point,
that determines the CEP as $(\mu_c, T_c)=(172.7, 159.2)$. Our numerical calculation
here for the critical temperature is in good agreement with the result
$T_c=\SI{154\pm 9}{MeV}$ obtained by lattice QCD~\cite{PhysRevD.85.054503}.
There is a long-standing debate, whether the chiral symmetry restoration
and the deconfinement transition have a correspondence
or not~\cite{Borsanyi2010}, whereas in our PNJL model the two phase
transitions coincide exactly and both are of first-order transitions.

\begin{figure}
 \includegraphics[width=235pt]{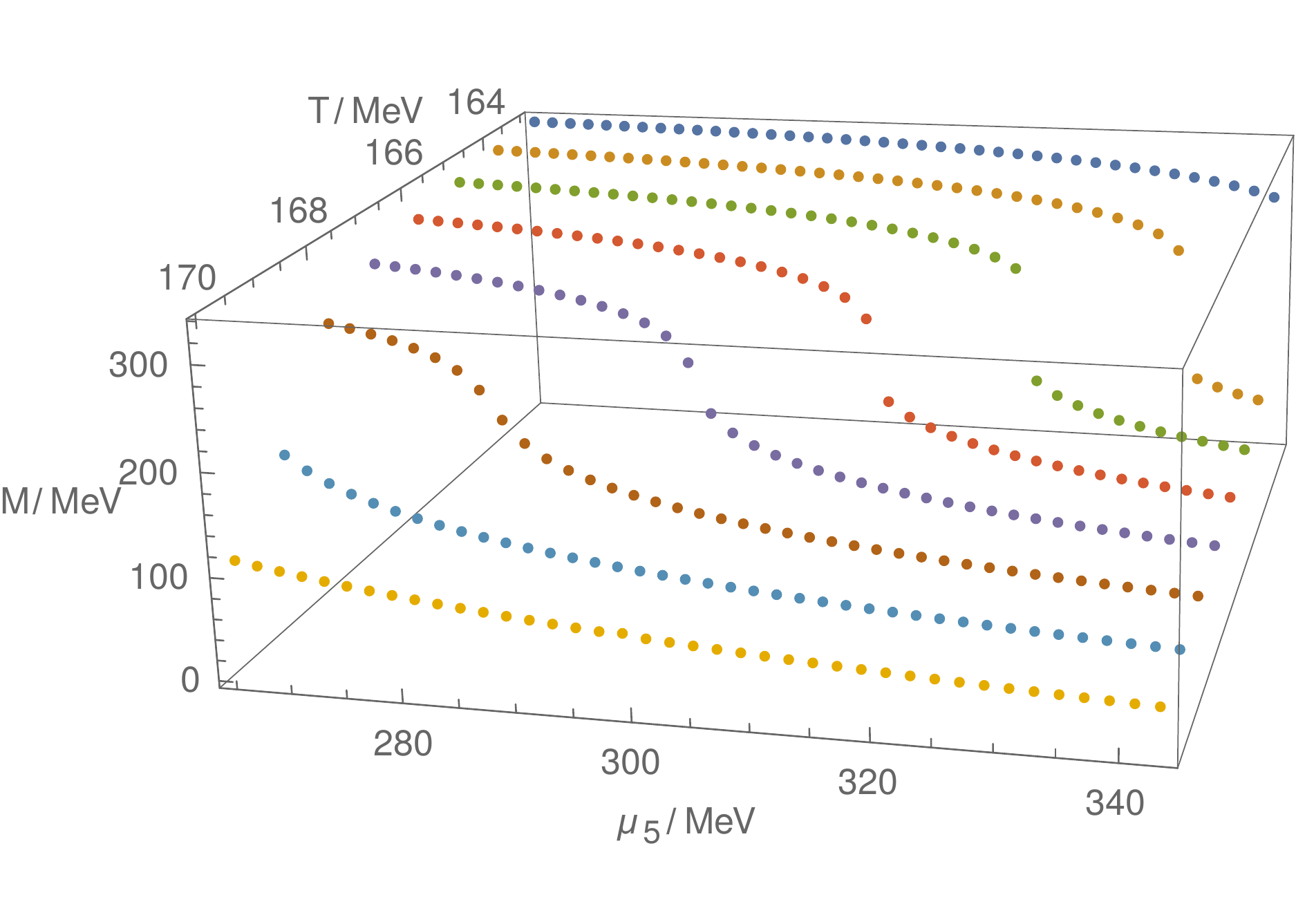}
 \includegraphics[width=235pt]{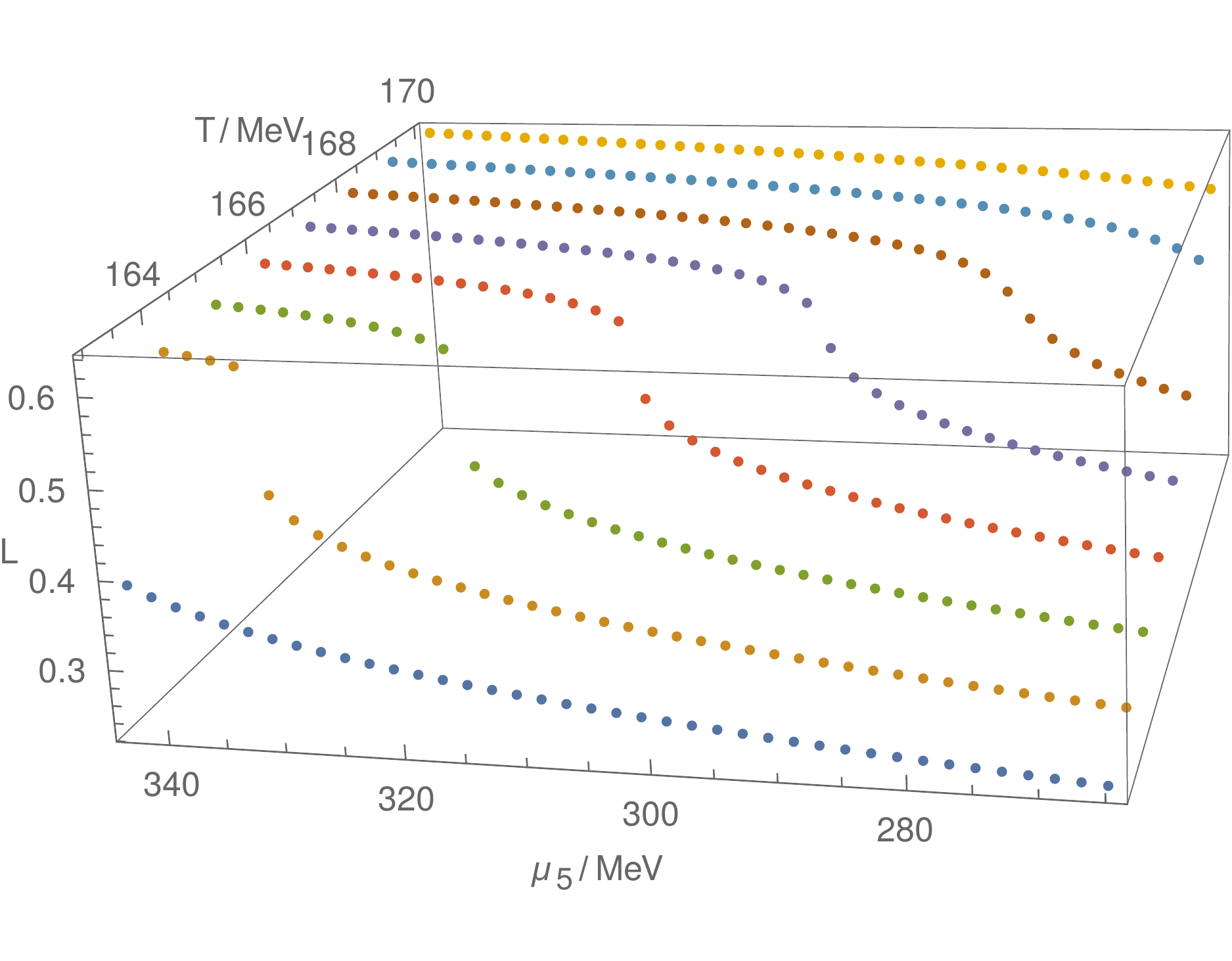}
 \caption{The 3D plot for the effective mass $M$ in $\mu_5$-$T$-$M$ space (the upper one)
  and the (3D) plot for the Polyakov loop $L$ in $\mu_5$-$T$-$L$ space (the lower one)
  near the $\text{CEP}_5$ $(\mu_{5c}, T_{5c})=(307.6, 166.1)$, where $\mu=0$.}
 \label{fig:mu-zero-3d}
\end{figure}

For $\mu=0$, we can also determine the location of $\text{CEP}_5$:
$(\mu_{5c}, T_{5c})=(307.6, 166.1)$, which is plotted in Fig.~\ref{fig:mu-zero-3d}.
It is interesting that the critical temperature is almost unchanged in
the continuation of CEP to $\text{CEP}_5$. In Ref.~\cite{PhysRevD.84.014011},
a novel idea on location of the CEP has been suggested
by using the relations $\mu_c/\mu_{5c}$. Namely when lattice simulations find $\text{CEP}_5$,
then it indicates a signal for the existence of the CEP in QCD.

As linear response of the physical system to some external field,
susceptibilities are widely used to study the phase transitions of
strongly interacting matter~\cite{Cui2015172}.
Here, let us consider three kinds of susceptibilities:
the vector susceptibility $\chi_{v}$, the axial-vector susceptibility
$\chi_{av}$, and the thermal susceptibility $\chi_T$.
They are defined as follows
\begin{equation}
 \chi_{v}=\frac{\partial\sigma}{\partial\mu},\quad
 \chi_{av}=\frac{\partial\sigma}{\partial\mu_5},\quad
 \chi_{T}=\frac{\partial\sigma}{\partial T}.
\end{equation}
All these susceptibilities are singular at the CEP or $\text{CEP}_5$ and
are continuous in the crossover region, so we may use this fact to
accurately identify the location of CEP and/or $\text{CEP}_5$.

\begin{figure}
 \includegraphics[width=235pt]{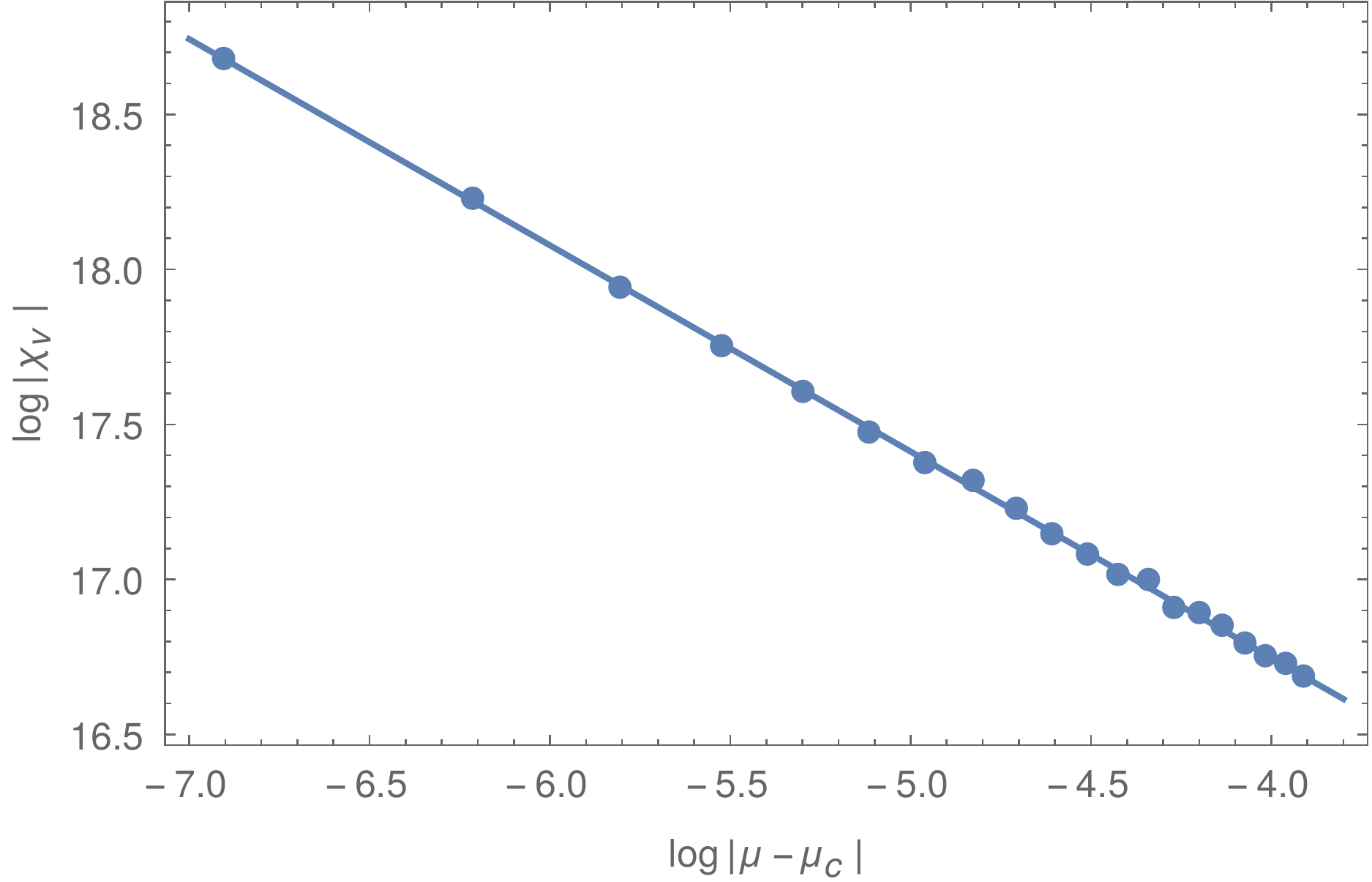}
 \caption{The linear fit with the logarithm of
  the vector-scalar susceptibility $\chi_{v}$ as a function of
  $\log|\mu-\mu_c|$ at the fixed temperature $T_c$ in the direction
  $\rightarrow$, where $\mu_5=0$. The critical exponent $\gamma_{v}$
  is calculated from the slope as $0.665\pm0.030$.}
 \label{fig:linear-fit}
\end{figure}

As well-known, a susceptibility in the vicinity of CEP or $\text{CEP}_5$
diverges in a power law with the so-called critical exponent $\gamma$.
These exponents are only dependent on the dimension of space and
the order parameter(s), but do not involve the details of microscopic dynamics.
Different systems in the same universality class all will share
the same critical behavior. Practically and for simplicity, we may choose a
specific direction, which is denoted by ``$\rightarrow$'', to calculate
the critical exponents by the path from lower $\mu$ or $\mu_5$ toward higher
$\mu_c$ or $\mu_{5c}$ with the temperature $T=T_c$ being fixed.
Using the linear logarithm fit, we obtain
\begin{equation}
 \begin{split}
  \log\chi&=-\gamma\log|T-T_c|+\text{const}.
 \end{split}
\end{equation}
The critical exponent of the vector-scalar susceptibility in the direction
``$\rightarrow$'' is calculated in Fig.~\ref{fig:linear-fit}.

Similarly, as necessary checks, we also calculate the critical exponent in the other directions
such as those denoted as ``$\leftarrow$'', ``$\uparrow$'', ``$\downarrow$'' with the path from
higher $\mu$ or $\mu_5$ toward $\mu_c$ or $\mu_{5c}$, the path from lower $T$ toward $T_c$,
and the path from higher $T$ toward $T_c$ respectively. The critical exponents of 
$\chi_{v}$ and $\chi_T$ in these four directions for $\mu_5=0$ are calculated 
and put in Table~\ref{tab:exponents-mu5-zero}; the critical exponents of $\chi_{av}$ 
and $\chi_T$ for $\mu=0$ are also calculated and put in Table~\ref{tab:exponents-mu-zero}.
The obtained critical exponents all agree with the predictions about the universality by 
mean field method~\cite{PhysRevD.67.014028, PhysRevD.77.096001}, and in Ref.~\cite{Lu2015},
more critical exponents are calculated in the NJL model.

For $\mu_5\neq 0$ and $\mu\neq 0$, CEP will naturally evolve into $\text{CEP}_5$.
The discontinuity of the effective mass and Polyakov loop parameter vanishes at the same CEP,
which means that the chiral symmetry restoration and the deconfinement transition coincide exactly.
The projection of the evolution of CEP onto the $\mu$-$\mu_5$ plane is
illustrated in Fig.~\ref{fig:cep-mu-mu5}, which also presents the results are in good
agreement with the results in Ref.~\cite{PhysRevD.84.014011}. Furthermore,
we verify that the nonzero chiral chemical potential does not change
the critical exponents, and that $\gamma_{v}$, $\gamma_{av}$ and $\gamma_T$
are all approximately equal to $2/3$. This implies that for the QCD phase diagram
our continuation of the CEP to a fictitious CEP
belonging to a phase diagram in the $\mu_5$-$T$ plane is reasonable.

\begin{table}
 \caption{Critical exponents for $\mu_5=0$,
  where the CEP is calculated as $(\mu_c, T_c)=(172.7, 159.2)$.}
 \label{tab:exponents-mu5-zero}
 \begin{tabular}{cccc}
  \hline
   Critical exponent & Path & Numerical result & MF exponent \\ \hline
   \multirow{4}{*}{$\gamma_{v}$}
     & $\rightarrow$ & $0.665\pm0.030$ & \multirow{4}{*}{$\dfrac{2}{3}$} \\
     & $\leftarrow$ & $0.636\pm0.059$ \\
     & $\uparrow$ & $0.679\pm0.061$ \\
     & $\downarrow$ & $0.659\pm0.085$ \\ \hline
   \multirow{4}{*}{$\gamma_{T}$}
     & $\rightarrow$ & $0.658\pm0.014$ & \multirow{4}{*}{$\dfrac{2}{3}$} \\
     & $\leftarrow$ & $0.664\pm0.010$ \\
     & $\uparrow$ & $0.671\pm0.012$ \\
     & $\downarrow$ & $0.664\pm0.017$ \\
   \hline
 \end{tabular}
\end{table}
\begin{table}
 \caption{Critical exponents for $\mu=0$,
  where the $\text{CEP}_5$ is calculated as $(\mu_{5c}, T_{5c})=(307.6, 166.1)$.}
 \label{tab:exponents-mu-zero}
 \begin{tabular}{cccc}
  \hline
   Critical exponent & Path & Numerical result & MF exponent \\ \hline
   \multirow{4}{*}{$\gamma_{av}$}
     & $\rightarrow$ & $0.697\pm0.068$ & \multirow{4}{*}{$\dfrac{2}{3}$} \\
     & $\leftarrow$ & $0.717\pm0.082$ \\
     & $\uparrow$ & $0.694\pm0.155$ \\
     & $\downarrow$ & $0.690\pm0.172$ \\ \hline
   \multirow{4}{*}{$\gamma_{T}$}
     & $\rightarrow$ & $0.668\pm0.042$ & \multirow{4}{*}{$\dfrac{2}{3}$} \\
     & $\leftarrow$ & $0.728\pm0.061$ \\
     & $\uparrow$ & $0.684\pm0.013$ \\
     & $\downarrow$ & $0.671\pm0.016$ \\
   \hline
 \end{tabular}
\end{table}

\section{\label{sec:volume}The finite-volume effects for  CEP and $\text{CEP}_5$}

Since the strongly interacting matter formed through
the energy deposition of colliding particles definitely has a finite volume,
it is very important to have a clear understanding of the
finite-volume effects to fully contemplate the thermodynamic phases,
which may be created in the experiments.
In Refs.~\cite{PhysRevD.91.051501, PhysRevC.91.041901},
the variations of susceptibilities with temperature
for different system sizes have been discussed, but here
we would like to focus on the finite-volume effects regarding
the CEP and $\text{CEP}_5$.

To incorporate the finite-volume effects,  methods such as the
Monte Carlo simulation~\cite{PhysRevD.81.114017} and the renormalization group
approach~\cite{PhysRevD.90.054012} may work, but we will make a lower momentum
cutoff $\lambda=\pi/R$ on the integration of the thermodynamic potential
density $\mathcal{V}$ of the extended PNJL model, where $R$ is the size of the concerned
system with a volume $\sim R^3$. Thus the thermodynamic potential density
Eq.~\eqref{eq:thermodynamic-potential} is rewritten as
\begin{equation}
 \label{eq:potential-momentum-cutoff}
 \begin{split}
  \mathcal{V}&=\mathcal{U}(L,L^{\dag},T)+G\sigma^2
   -N_cN_f\sum_{s=\pm 1}\int_{\lambda}^{\Lambda}
   \frac{\mathrm{d}^3\bm{p}}{(2\pi)^3}\,\omega_s \\
  &\qquad -N_f\sum_{s=\pm 1}\int_{\lambda}^{\infty}
   \frac{\mathrm{d}^3\bm{p}}{(2\pi)^3}\,T\log(F_{+}F_{-})\,,
 \end{split}
\end{equation}
here, several simplifications are made. The most important one is
the infinite sum of Fourier series is replaced by an integration over
a continuous variation of momentum with the lower cutoff. We also neglect
the surface and curvature effects, and the slight changes due to introducing
the lower cutoff for the other parameters, such as $\sigma$
and the relevant ones in $\mathcal{U}(L,L^{\dag},T)$.
The same simplifications to consider the finite-volume effects
can also be found in Ref.~\cite{PhysRevD.87.054009, PhysRevD.91.051501}.
Obviously, when $\lambda=0$ is taken (corresponding to $R=\infty$) in
Eq.~\eqref{eq:potential-momentum-cutoff}, all reduces to the case of infinite volume.

\begin{figure}
 \includegraphics[width=235pt]{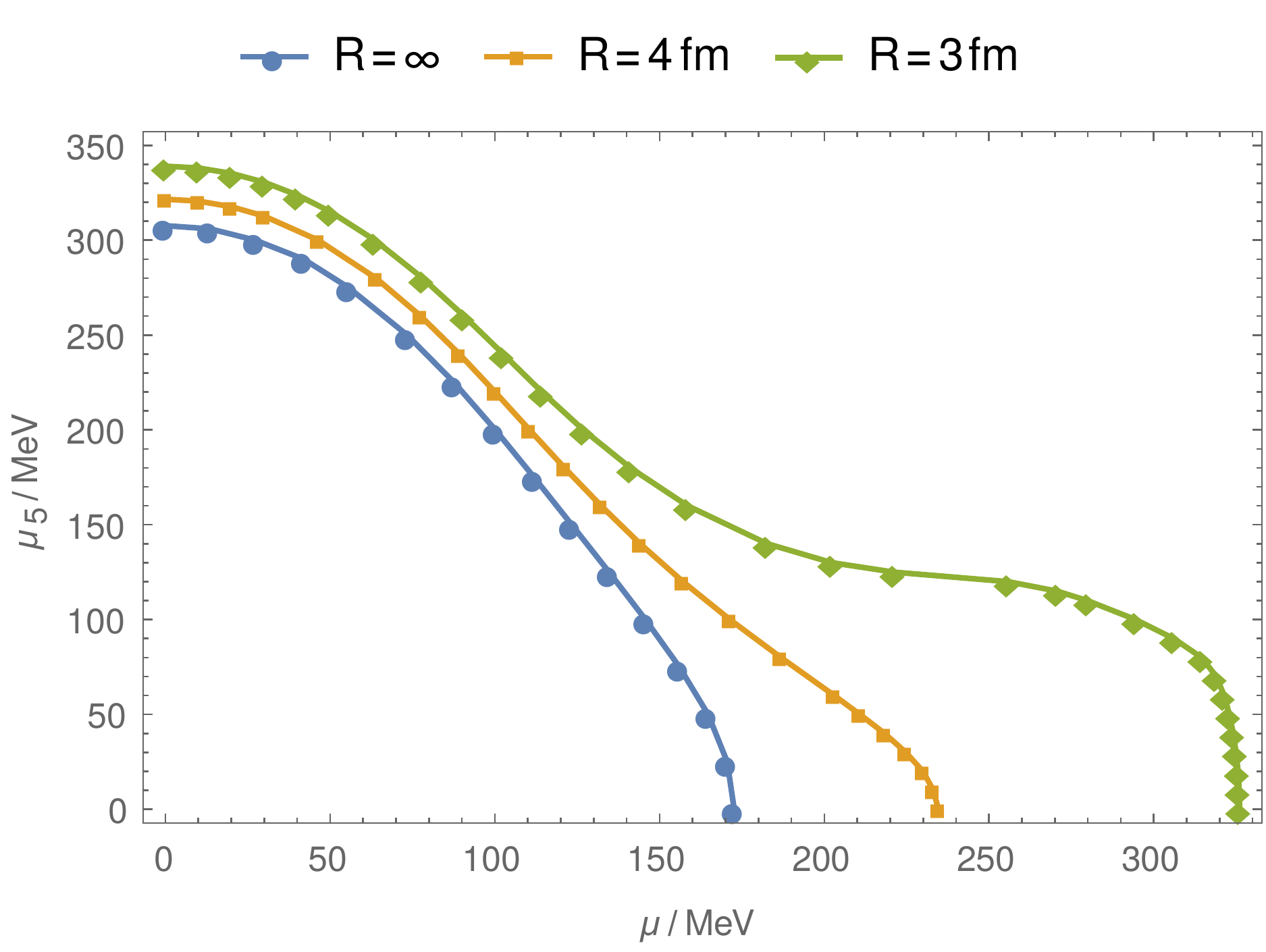}
 \caption{Projection of the evolution of CEP on the $\mu$-$\mu_5$ plane
  for different system sizes: $R=\infty$, \SI{4}{fm}, and \SI{3}{fm}.}
 \label{fig:cep-mu-mu5}
\end{figure}

Following the same procedure as that in the previous section,
we find the location of CEP or $\text{CEP}_5$ for various system sizes,
and calculate out the corresponding critical exponents, and it has been found that
the system size $R$ does not affect the critical exponents.
In Fig.~\ref{fig:cep-mu-mu5}, we plot the projection of the evolution of
CEP to $\text{CEP}_5$ on the $\mu$-$\mu_5$ plane for various system sizes:
$R=\infty$, \SI{4}{fm} and \SI{3}{fm} respectively. One can see clearly that
the finite-volume effects become more and more manifest as $R$ decreases,
and are more important for lower $\mu_5$. We have also found that the CEP
vanishes at a $R_{\text{min}}$, whose precise value is estimated to be about $\SI{2.1}{fm}$.
Namely when $R<R_{\text{min}}$, the whole phase diagram becomes a crossover.
Our numerical results in Table~\ref{tab:cep-cep5-relations} also show that the ratios
$\mu_c/\mu_{5c}$ and $T_c/T_{5c}$ are significantly affected by system sizes.
If we use the idea proposed in Ref.~\cite{PhysRevD.84.014011} to find the location of CEP,
the finite-volume effects on these ratios should be taken into account.

\begin{figure}
 \includegraphics[width=235pt]{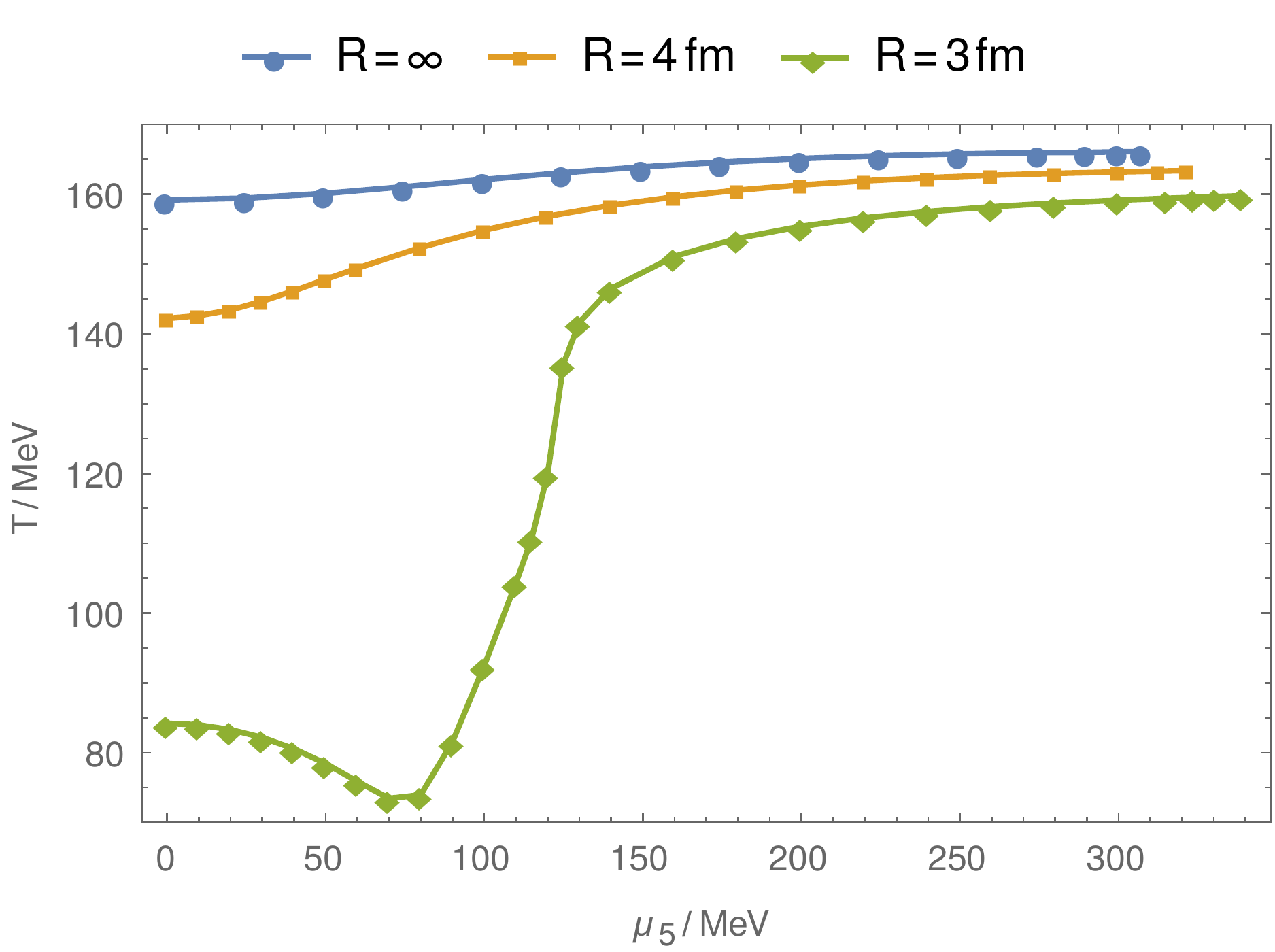}
 \caption{Projection of the evolution of CEP on the $\mu_5$-$T$ plane
  for different system sizes: $R=\infty$, \SI{4}{fm}, and \SI{3}{fm}.}
 \label{fig:cep-mu5-t}
\end{figure}

In Fig.~\ref{fig:cep-mu5-t}, we plot the projection of the evolution of
CEP on the $\mu_5$-$T$ plane for different system sizes.
From the plot one can see that the relations between $T_c$ and $\mu_5$ are intriguing:
when $R$ is large, $T_c$ increases slowly with $\mu_5$; when $R$ is small,
$T_c$ decreases first and then increases with $\mu_5$. In the case of large volumes, 
our results are qualitatively consistent with the results obtained
within the framework of Dyson-Schwinger equations~\cite{PhysRevD.91.056003}
and the lattice simulation~\cite{PhysRevD.93.034509}. However, we should
point out that the other studies with different models or methods~\cite{PhysRevD.81.114031,
PhysRevD.83.105008, PhysRevD.88.054009, PhysRevD.90.125025}
may have given opposite results in $T_c$ decreasing with $\mu_5$.
Whereas our calculations show that the dependence of $T_c$ on $\mu_5$
is changed by the system sizes, which means the finite-volume effects
are very important in studying the phase transitions of effective models in QCD.

\begin{table}
 \caption{Numerical relations between
  CEP and $\text{CEP}_5$ for different system sizes.}
 \label{tab:cep-cep5-relations}
 \begin{tabular}{cccc}
  \hline
   $R$ & $(\mu_c, T_c)$ & $(\mu_{5c}, T_{5c})$ & $(\mu_c/\mu_{5c}, T_c/T_{5c})$ \\ \hline
   $\infty$ & $(172.7, 159.2)$ & $(307.6, 166.1)$ & $(0.561, 0.958)$ \\
   \SI{5}{fm} & $(194.5, 153.6)$ & $(314.9, 164.7)$ & $(0.618, 0.933)$ \\
   \SI{4}{fm} & $(234.6, 142.2)$ & $(321.6, 163.4)$ & $(0.729, 0.870)$ \\
   \SI{3}{fm} & $(326.0, 84.2)$ & $(339.0, 159.8)$ & $(0.962, 0.527)$ \\
  \hline
 \end{tabular}
\end{table}

\begin{figure}
 \includegraphics[width=235pt]{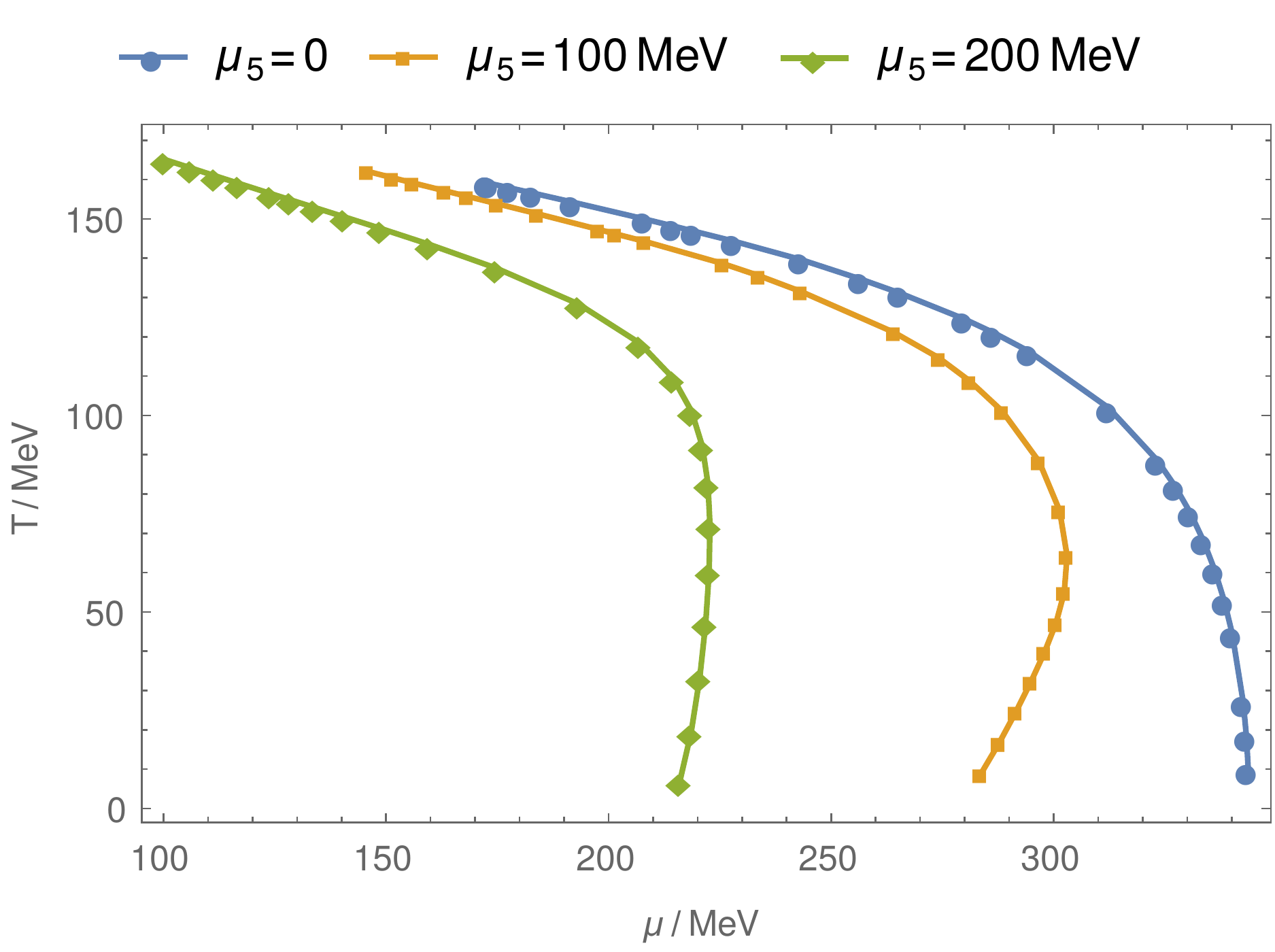}
 \caption{Projection of the evolution of CEP on the $\mu$-$T$ plane for
  fixed chiral chemical potential: $\mu_5=0$, \SI{100}{MeV}, and \SI{200}{MeV},
  varying in the system sizes from $R=\infty$ with the highest critical
  temperature to $R_{\text{min}}$ with the lowest critical temperature.
  The corresponding values of $R_{\text{min}}$ are estimated around
  \SI{2.1}{fm}, \SI{2.0}{fm}, and \SI{1.6}{fm} respectively.}
 \label{fig:cep-mu5-fixed}
\end{figure}

Another interest point is to see the shift of CEP in the phase diagram
with respect to the system size. In Fig.~\ref{fig:cep-mu5-fixed}, we plot the projection 
of the evolution of CEP on the $\mu$-$T$ plane for fixed chiral chemical potential:
$\mu_5=0$, \SI{100}{MeV}, and \SI{200}{MeV} respectively, varying the system sizes
from $R=\infty$ with the highest critical temperature to $R_{\text{min}}$ with
the lowest critical temperature. The corresponding values of $R_{\text{min}}$
are estimated around \SI{2.1}{fm}, \SI{2.0}{fm}, and \SI{1.6}{fm} respectively.
It is interesting to note that the relation between critical temperature and
the system size may be nonmonotonic for some values of the chiral chemical potential
such as $\mu_5=\SI{200}{MeV}$.

Our results on the finite-volume effects have significant implications
for heavy-ion collision experiments, because the strong interacting matter
formed in a heavy-ion collision is finite in volume, and its size depends
on the size of the colliding nuclei, the collision center of mass energy $\sqrt{s}$,
and the centrality of the collision. In our concerned case, $\sqrt{s}$
and the centrality of collisions relate to the temperature $T$ and the chiral
chemical potential $\mu_5$ when the colliding nuclei are chosen.
It is expected that our results will provide some guidance for the experiments
aiming at the search of the possible critical end point.
Although there have been many efforts to estimate
the size of the strong interacting matter formed by heavy-ion
collisions ~\cite{Habich2015, PhysRevC.91.054915, Bozek2013250,
PhysRevC.87.064906}, no general consensus have been reached.
In Ref.~\cite{PhysRevC.85.044901}, the system volume for Pb-Pb collisions
with $\sqrt{s}$ in the range of $62.4$ to \SI{2760}{GeV}
has been estimated in the range to vary from $50$ to \SI{250}{fm^3},
corresponding to a system size from $3$ to \SI{6}{fm}.
Given that these are the volumes at the time of freeze out,
one may expect an even smaller system size at the initial
equilibration time~\cite{Bozek2013250, PhysRevC.87.064906}.
Note that our estimate adopts several approximations,
then to test of our numerical results, possible uncertainties should
be taken into account.

\section{\label{sec:conclusions}Summary and Conclusion}

To summarize, we have studied the chiral symmetry restoration and
the deconfinement transition of the phase diagram of QCD by using the
extended PNJL model in which chemical and chiral chemical potentials
$\mu$ and $\mu_5$ are introduced. Our results show that the discontinuity of
the effective mass and Polyakov loop parameter always vanishes at the same CEP,
and the chiral symmetry restoration and the deconfinement, the two transitions,
coincide exactly. Three kinds of susceptibilities are defined and the corresponding
critical exponents are calculated. All the critical exponents are approximately equal to $2/3$.
We also show that the critical exponents do not change even artificially introducing 
a chiral chemical potential $\mu_5$. This implies that the continuation for CEP of
the QCD phase diagram to a fictitious CEP belonging to a phase diagram
in the $\mu_5$-$T$ plane is also meaningful.

Finally, it is found that the finite-volume effects become more and more
manifest as $R$, the radius of the size for the considered finite-volume,
decreases and are more important for a smaller $\mu_5$, by introducing a
lower momentum cutoff in the integration in $\mathcal{V}$ as
Eq.~\eqref{eq:potential-momentum-cutoff} to investigate the finite-volume
effects in the PNJL model. Numerical results show that the ratios
$\mu_c/\mu_{5c}$ and $T_c/T_{5c}$ are significantly affected by
the system sizes with radius $R$. When $R$ is comparatively large,
$T_c$ increases slowly with $\mu_5$; when $R$ is comparatively small,
$T_c$ decreases first and then increases with $\mu_5$. For a fixed $\mu_5$,
we can also determine a $R_{\text{min}}$ such that the CEP
vanishes when $R<R_{\text{min}}$, and the whole phase diagram becomes a crossover.
The corresponding values of $R_{\text{min}}$ for $\mu_5=0$, \SI{100}{MeV}
and \SI{200}{MeV} are estimated around \SI{2.1}{fm}, \SI{2.0}{fm}
and \SI{1.6}{fm} respectively. Therefore, the present results on
the finite-volume effects indicate that when investigating the CEP of 
QCD phases experimentally, the volume of strong interacting matter in 
the heavy-ion collision cannot be small, otherwise the result on CEP would be flexible.

\begin{acknowledgements}
This work is supported in part by the National Natural Science Foundation of China
under the Grants No. 11275243, No. 11535002, No. 11447601, No. 11275097, No. 11475085
and No. 11535005, the China Postdoctoral Science Foundation under the Grant No. 2015M581765,
and the Jiangsu Planned Projects for Postdoctoral Research Funds under the Grant No. 1402006C.
\end{acknowledgements}

\end{document}